\documentclass[aps,pre,reprint,showkeys,showpacs,amsfonts,amsmath,amssymb,superscriptaddress,floatfix]{revtex4-1}

\usepackage[ansinew]{inputenc}
\usepackage{graphicx}
\usepackage[english]{babel}
\usepackage{color}

\newcommand{\op}[1]{\hat{\mathrm{#1}}}

\DeclareMathOperator*{\SumInt}{%
\mathchoice%
  {\ooalign{$\displaystyle\sum$\cr\hidewidth$\displaystyle\int$\hidewidth\cr}}
  {\ooalign{\raisebox{.14\height}{\scalebox{.7}{$\textstyle\sum$}}\cr\hidewidth$\textstyle\int$\hidewidth\cr}}
  {\ooalign{\raisebox{.2\height}{\scalebox{.6}{$\scriptstyle\sum$}}\cr$\scriptstyle\int$\cr}}
  {\ooalign{\raisebox{.2\height}{\scalebox{.6}{$\scriptstyle\sum$}}\cr$\scriptstyle\int$\cr}}
}

\begin{document}

\title{Targeting Eigenstates using a Decoherence based Nonlinear Schr\"{o}dinger Equation}

\author{O. Furtmaier}
\email{oliverfu@ethz.ch}
\affiliation{ETH Z\"{u}rich, Computational Physics for Engineering Materials, Institute for Building Materials, Schafmattstrasse 6, HIF, 8093 Z\"{u}rich, Switzerland}
\author{M. Mendoza}
\email{mmendoza@ethz.ch}
\affiliation{ETH Z\"{u}rich, Computational Physics for Engineering Materials, Institute for Building Materials, Schafmattstrasse 6, HIF, 8093 Z\"{u}rich, Switzerland}

\begin{abstract}
Inspired by the idea of mimicking the measurement on a quantum system through a decoherence process to target specific eigenstates based on Born's law, 
i.e. the hiearchy of probabilities instead of the hierarchy of eigenvalues
, we transform a Lindblad equation for the reduced density operator into a nonlinear Schr\"{o}dinger equation 
to obtain a computationally feasible simulation of the decoherent dynamics in the open quantum system. This gives the opportunity to target the eigenstates which have the largest 
$L^2$ overlap with an initial superposition state and hence more flexibility in the selection criteria. One can use this feature for instance to approximate eigenstates with certain localization or symmetry properties.
As an application of the theory we discuss \textit{eigenstate towing}, which relies on the perturbation theory to follow the progression of an arbitrary subset of 
eigenstates along a sum of perturbation operators with the intention to explore for example the effect of interactions on these eigenstates.
The easily parallelizable numerical method shows an exponential convergence and its computational costs scale linear for sparse matrix representations of the involved Hermitian operators.
%Symmetries of the problem can be incorporated either in the initial state of the dynamics or explicitly using the symmetry operators in the evolution equation.
\end{abstract}


\maketitle

\section{Introduction}
\label{intro}
What makes eigenstates and -values of Hermitian operators interesting for physics? 
A postulate of the early days quantum theory says that the measurement intervention ``instantaneously'' causes the state of an isolated quantum mechanical system, 
which is a unit vector in a countably infinite dimensional Hilbert space, to collapse to one random eigenstate of the Hermitian operator that represents the measured observable, 
such that the probabilities are given by Born's rule \cite{born26,wheeler83}.
With \textit{collpase}, we mean the sudden reduction of the superposition of eigenstates to only one eigenstate of the observable which is then occupied by the quantum system.
Nowadays, many scientists believe that environment-induced decoherence offers a more fundamental explanation for this phenomenon and reconciles the deterministic, unitary, 
continuous time evolution of the linear Schr\"{o}dinger equation with the non-deterministic, non-unitary, discontinuous reduction of the wave function in the collapse 
\cite{zurek91,zurek14}. 
Hereby, the local interaction between the measurement apparatus and the open quantum system, which together evolve according to the linear Schr\"{o}dinger equation, 
generates entanglement that (usually irreversibly) spreads the coherence of the quantum system into the environment, or measurement apparatus, 
and will finally result in a complete loss of quantum phase information, i.e. the system becomes a classical mixture of preferred states that satisfies Born's probability law \cite{zurek03b,schloss05b}. 
The (ein-)selection of these (pointer) states is induced by the system-environment interaction which measures certain observables of the system 
and hence leads to the preference of eigenstates of the corresponding Hermitian operators \cite{paz99}.

A popular way of modeling the process is an open quantum system with a non-unitary evolution that deflates interference, 
i.e. reduces the off-diagonal elements of the reduced density operator in the pointer basis \cite{schloss05a,zurek05}.
As indicated in Ref. \cite{weinberg16}, a master equation in Lindblad form \cite{lindblad76} can be used to achieve this effect.
The reduced density operator, which is formally obtained by averaging over the environment's degrees of freedom, contains all information
that is accessible via measurements on the quantum system alone \cite{neumann32}. 
There are other, more general, starting points for decoherence dynamics of the reduced density operator that explicitly ignore the environmental degrees of freedom, 
such as integro-differential equations which are not local in time (memory effects) \cite{nakajima58}, non-Markovian master equations \cite{stamp00} 
or the Born-Markov master equation \cite{breuer02}. However, the Lindblad equation offers an intuitive and easy representation of the measurement process in the weak coupling limit 
\cite{weinberg16}.

Based on these insights, one can use decoherent dynamics to approximate specific eigenstates of a finite set of dimensionless commutable Hermitian operators with discrete spectra 
$\{\hat{\mathrm{O}}_j\}_{j\in\mathcal{J}}$. One can imagine this as measurements on the quantum system by simultaneously and continuously monitoring the observables 
$\mathrm{O}_j$. To achieve this with a reasonable computational effort, we have formulated a deterministic, nonlinear Schr\"{o}dinger equation, similar to the equation in 
Ref. \cite{gisin95}, such that the dark states of the open quantum system are given by the eigenstates of $\{\hat{\mathrm{O}}_j\}_{j\in\mathcal{J}}$. 
By dark states we mean the states which are unaffected by the coupling to an environment, see Ref. \cite{diehl08}.
The equation is derived from a purely decoherent Lindblad equation for the reduced density operator, i.e. we work in the quantum-measurement limit \cite{schloss07}. 
A different approach for the problem could have been the use of quantum trajectories \cite{gisin84,diosi88}, which also reduces the required computational resources compared to the 
master equation for the reduced density operator, but still needs the simulation of many trajectories to approximate the evolution according to the 
master equation with a reasonable accuracy. With our deterministic equation we sacrifice the exact compliance with Born's law for the gain of computational speed.
The algorithm's computational costs scale linear for sparse matrix representations of the involved Hermitian operators and one perturbation step.
In addition, there are two simple implementations which can make use of symmetry induced subspaces. One option is to use a symmetry adapted starting vector to initialize 
the dynamics. The other option is to include the symmetry operator explicitly in the dynamics. Nice examples are the initialization of fermionic or bosonic wave functions, which are
fully (anti-)symmetric with respect to the exchange of particles, that will result in fully (anti-)symmetric eigentstates through the dynamics if the self-adjoint operators commute
with this symmetry operation, as shown in appendix \ref{SymInh}.

There are other successful methods, \cite{lanczos50,bathe76,hammond94,feng96,lehto07}, which allow the computation of the low-lying eigenvalue states. 
However, one might be interested in states that possess other characteristics such as higher energy eigenstates (orbitals) of the Kohn-Sham equations \cite{aich05}, 
states with certain localization properties, the influence of interactions on a certain subset of eigenstates in the spectrum or excited-state quantum phase transitions \cite{caprio08}.
By using the ``folded spectrum'' method \cite{wang94}, i.e. folding the spectrum of each operator around a given reference eigenvalue $\epsilon^{\text{ref}}_j$ 
and using the square of the shifted self-adjoint operator $\left(\hat{\mathrm{O}}_j-\epsilon^{\text{ref}}_j\hat{\mathbb{I}}\right)^2$, 
the methods in Refs. \cite{lanczos50,bathe76,hammond94,feng96} would converge to the eigenstates whose eigenvalues are closest to the reference.
Furthermore, as mentioned in chapter six of Ref. \cite{hammond94}, there are extensions which facilitate the approximation of excited-states and make use 
of symmetries to improve the convergence.
Nevertheless, all the methods mentioned so far do not allow the computation of eigenstates based on the $L^2$ overlap with an arbitrary quantum state and 
mainly focus on the eigenvalue and symmetry properties as selection criteria. 
For example, the $L^2$ overlap of an eigenstate with a given quantum state may imply among other things certain localization characteristics, such as edge or surface states.
The numerical procedure introduced in Ref. \cite{tackett02} allows the ``targeting of specific eigenvectors using arbitrary physical properties as selection criteria''. 
An advantage of this method is that it very well differentiates nearly degenerate states because using this procedure to target a state with a reference value 
$\epsilon^{\text{ref}}_j$ the respective eigenstate $|\psi_i\rangle$ with 
$\hat{\mathrm{O}}_j|\psi_i\rangle = \epsilon_{j,i}|\psi_i\rangle$ has an eigenvalue $1/(\epsilon_{j,i}-\epsilon^{\text{ref}}_j)$. 
In addition, it works with the linear instead of the quadratic operator which results in a smaller condition number and 
hence may reduce the numerical difficulty for solving. The disadvantage of this approach is that the employed Jacobi-Davidson method has only cubic convergence and that one needs
to store the vectors in the search space, which increases the memory consumption compared to just storing a single vector that is going to approximate the desired eigenvector.

In summary, we are proposing a decoherence based approach, modeled by a Lindblad equation, which makes use of Born's law. In section \ref{deriv} 
we will show that the only stable equilibria of the dynamics are given by the eigenstates and demonstrate the exponential convergence with 
$\text{exp}\left[-\sum_{j\in\mathcal{J}}(\epsilon_{j,i}-\epsilon_{j,i'})^2 \tau\right]$ to one of the eigenstates contained in the input superposition, 
which is usually the one with the highest probability. This allows us to target eigenstates that maximize the $L^2$ overlap with a predefined function. 
Furthermore, in section \ref{eigen_tow} an application of the dynamics, that we would describe as \textit{eigenstate towing}, is discussed.
The goal of the method is to follow the progression of an abitrary subset of eigenstates along a perturbation strength increase. 
It makes use of perturbation theory \cite{schroed26}, which ensures the collapse of the unperturbed eigenstate into the same eigenstate after an infinitesimal perturbation. 
This enables us to avoid any communication between the computations for different eigenstates, because each one will converge independent of the others
and a parallel implementation is achievable. It should be mentioned that also other methods in quantum chemistry can approximate lower eigenstates in parallel, 
see Refs. \cite{sergio16,goedecker99}. However, the method in Ref. \cite{sergio16} relies on an information exchange between the different computations to avoid the convergence
to eigenstates with smaller eigenvalues.
In section \ref{num_impl} we propose the (semi-)implicit Crank-Nicholson method for a specific numerical implementation, 
that is used in section \ref{ESQPT} to analyze an example of an excited-state quantum phase transition in the Jaynes-Cummings model \cite{jaynes63,tavis68,arias11}.
This problem also serves to compare our algorithm, in section \ref{benchmarking}, with one of the fastest LAPACK algorithms for determining eigenvectors and -values of tridiagonal symmetric matrices \cite{demmel08}, 
the \textit{multiple relatively robust representations} (MRRR) algorithm \cite{dhillon97,dhillon06}.

%%%%%%%%%%%%%%%%%%%%%%%%%%%%%%%%%%%%%%%%%%%%%%%%%%%%%%%%%%%%%%%%
%%%%%%%%%%%%%%%%%%%%%%%%%%%%%%%%%%%%%%%%%%%%%%%%%%%%%%%%%%%%%%%%
\section{Derivation \& Properties of Nonlinear Schr\"{o}dinger Equation}
\label{deriv}
We assume a finite set of commutable self-adjoint operators $\{\hat{\mathrm{O}}_j\}_{j\in\mathcal{J}}$ with discrete spectra on the Hilbert space $\mathcal{H}$.
From the spectral theorem \cite{sakurai14} we know there exists a common complete eigenbasis $\{|\psi_{\vec{a}}\rangle\}_{\vec{a}\in\mathcal{A}}$ for this set that spans the whole Hilbert space, i.e.
$\forall \mathcal{I}\subseteq\mathcal{J}$
\begin{equation}
 \left(\prod_{i\in\mathcal{I}}\hat{\mathrm{O}}_i\right)|\psi_{\vec{a}}\rangle = \left(\prod_{i\in\mathcal{I}}a_i\right)|\psi_{\vec{a}}\rangle \text{ ,}
\end{equation}
and $\forall |\psi\rangle\in\mathcal{H}$
\begin{equation}
  |\phi\rangle = \sum_{\vec{a}\in\mathcal{A}} b_{\vec{a}}|\psi_{\vec{a}}\rangle \text{ , with } b_{\vec{a}}\in\mathbb{C}\text{ .}
\end{equation}
The operators are Abelian such that a common eigenbasis for all of them exists. The usage of more than one Hermitian operator serves the purpose of fine-tuning 
the targeting of a specific eigenstate in the basis by lifting potential degeneracies in the spectrum.
As an example one can think of the bound eigenstates of the hydrogen atom which can be labeled by their eigenenergy $\langle\hat{\mathrm{H}}\rangle$,
the total angular momentum $\langle|\hat{\vec{\mathrm{J}}}|^2\rangle$ and the $z$-component $\langle \hat{\mathrm{J}}_z\rangle$.
In an attempt to model the quantum mechanical measurement process and mimic the collapse of the wave function, in analogy to the ideas expressed in Ref. \cite{weinberg16}, 
we employ the Lindblad equation for the density operator
\begin{align}
 \partial_t \hat{\rho}(t) &= \sum_{j\in\mathcal{J}}\left[[\hat{\mathrm{O}}_j,\hat{\rho}(t)],\hat{\mathrm{O}}_j\right] \text{ ,}\\
 &= -\sum_{j\in\mathcal{J}} \left(\hat{\mathrm{O}}_j^2 \hat{\rho}(t) + \hat{\rho}(t)\hat{\mathrm{O}}_j^2 - 2\hat{\mathrm{O}}_j\hat{\rho}(t)\hat{\mathrm{O}}_j\right)\text{ ,}\label{LindDyn}
\end{align}
where we used the observables as Lindblad operators. 
One can interpret this dynamics as a continuous measurement on a quantum system which was
at the beginning of the process in the pure state $\hat{\rho}(0) = |\phi\rangle\langle\phi|$. However, we would like to point out that it is still unknown how the measurement
process in quantum mechanics exactly works, giving rise to different interpretations of quantum mechanics, see Ref. \cite{schloss13}, and that here we are using a decoherence 
based approach to model it.
To better understand the evolution in Eq. \eqref{LindDyn}, 
we look at the temporal change of the coefficients of the density operator in the eigenbasis representation of our operator set, which reads
\begin{align}
 \hat{\rho}(t) &= \sum_{\vec{a},\vec{a}'\in\mathcal{A}} c_{\vec{a},\vec{a}'}(t) |\psi_{\vec{a}}\rangle\langle\psi_{\vec{a}'}|\text{ ,}\\
 \partial_t c_{\vec{a},\vec{a}'} &= -|\vec{a}-\vec{a}'|^2 c_{\vec{a},\vec{a}'}(t)\text{ ,}\\
 c_{\vec{a},\vec{a}'}(0) &= b_{\vec{a}}b_{\vec{a}'}^*\text{ .}
\end{align}
The dynamics are purely decoherent, i.e. that 
expectation values with respect to the operators $\hat{\mathrm{O}}_j$ are unchanged and only off-diagonal elements in the chosen representation decay exponentially 
which in the infinite limit results in a fully classical mixture
\begin{align}
\hat{\rho}(t) &= \sum_{\vec{a},\vec{a}'\in\mathcal{A}}e^{-|\vec{a}-\vec{a}'|^2 t} b_{\vec{a}}b_{\vec{a}'}^* |\psi_{\vec{a}}\rangle\langle\psi_{\vec{a}'}| \text{ ,}\\
 \lim\limits_{t\rightarrow \infty} \hat{\rho}(t) &= \sum_{\vec{a}\in\mathcal{A}} |b_{\vec{a}}|^2 |\psi_{\vec{a}}\rangle\langle \psi_{\vec{a}}| \text{ ,}
\end{align}
which means that all the quantum correlations have vanished and there is only a statistical ensemble of quantum states, 
i.e. the system is in a mixed state with classical probabilities.
To reduce the dimensionality of the problem, similar to Refs. \cite{diosi86,gisin95}, we project the density operator onto the initial quantum state of the system, 
i.e. $|\phi(t)\rangle \equiv \hat{\rho}(t)|\phi\rangle$ where $|\phi\rangle\langle\phi| \equiv \hat{\rho}(0)$ is the initial, 
normalized, pure state of the quantum system. The resulting dynamics is 
\begin{equation}
 \partial_t |\phi(t)\rangle = \dot{\hat{\rho}}(t)|\phi\rangle \text{ ,}
\end{equation}
and the first order approximation looks like
\begin{align}
 |\phi(t+\delta t)\rangle &\approx |\phi(t)\rangle + \delta t \sum_{j\in\mathcal{J}}\left[\left[\hat{\mathrm{O}}_j,\hat{\rho}(t)\right],\hat{\mathrm{O}}_j\right]|\phi\rangle\text{ ,}\\
 &= \left[\mathbb{I}-\delta t \sum_{j\in\mathcal{J}}\hat{\mathrm{O}}_j^2\right]|\phi(t)\rangle\notag\\
 &+ \delta t \sum_{j\in\mathcal{J}}\left[2\hat{\mathrm{O}}_j\hat{\rho}(t)\hat{\mathrm{O}}_j-\hat{\rho}(t)\hat{\mathrm{O}}_j^2\right]|\phi\rangle\text{ .}
\end{align}
The issue with this approach is that it requires the knowledge of the current density operator $\hat{\rho}(t)$ to evolve the wave function by an infinitesimal time $\delta t$. 
Nevertheless, one can advance the initial state, since $\hat{\rho}(0)=|\phi\rangle\langle\phi|$ is known from the starting condition. We find
\begin{equation}
 \frac{|\phi(\delta t)\rangle - |\phi\rangle}{\delta t} = \sum_{j\in\mathcal{J}}\left[\left[\hat{\mathrm{O}}_j,|\phi\rangle\langle\phi|\right],\hat{\mathrm{O}}_j\right]|\phi\rangle\text{ .}
\end{equation}
Following this line of thought, we write down the \textit{measurement dynamics} by making the replacements 
\begin{align}
 \hat{\rho}(t) &\rightarrow |\phi(t)\rangle\rangle\langle\phi(t)|\text{ ,}\\
 |\phi\rangle &\rightarrow |\phi(t)\rangle\text{ ,}\\
 |\phi(0)\rangle &= |\phi\rangle \text{ ,}
\end{align}
in the previous equation, which results in
\begin{equation}
 \partial_t\left(|\phi(t)\rangle\langle\phi(t)|\phi(t)\rangle\right)=\sum_{j\in\mathcal{J}}\left[[\hat{\mathrm{O}}_j,|\phi(t)\rangle\langle\phi(t)|],\hat{\mathrm{O}}_j\right]|\phi(t)\rangle\text{ .}\label{ProjLind0}
\end{equation}
The nonlinearity in the equation follows from the replacement $\hat{\rho}(t) = |\phi(t)\rangle\langle\phi(t)|$ and the projection on $|\phi(t)\rangle$.
We would like to point out that this equation does not possess all the properties of Eq. \eqref{LindDyn}. 
Nevertheless, it allows for the reduction of the wave function to a single eigenstate in the spectrum, which will be shown subsequently and is the main purpose of it.
If we rewrite the equation as
\begin{align}
 \partial_t |\phi(t)\rangle &= \left[\sum_{j\in\mathcal{J}}\hat{\mathrm{B}}^{(\phi_t)}_j-\frac{\dot n(t)}{n(t)}\hat{\mathbb{I}}\right]|\phi(t)\rangle\text{ ,}\\
 \hat{\mathrm{B}}^{(\phi_t)}_j &\equiv 2\mathbb{E}_j^{(\phi_t,1)}\hat{\mathrm{O}}_j-\left(\hat{\mathrm{O}}_j\right)^2 - \mathbb{E}_j^{(\phi_t,2)}\hat{\mathbb{I}}\text{ ,}\label{B_OP}\\
 \mathbb{E}_j^{(\phi_t,k)} &\equiv \frac{1}{n(t)}\left\langle\phi(t)\left|\left(\hat{\mathrm{O}}_j\right)^k\right|\phi(t)\right\rangle\text{ ,}\\
 n(t) &\equiv \langle\phi(t)|\phi(t)\rangle\text{ ,}
\end{align}
where $\mathbb{E}_j^{(\phi_t,k)}$ stands for the expectation value of the observable $\left(\hat{\mathrm{O}}_j\right)^k$ in the state $|\phi(t)\rangle$,
and if we calculate the scalar product with $\langle\phi(t)|$ as well as the scalar product of $|\phi(t)\rangle$ with the complex conjugate of the 
previous equation we will find
\begin{align}
 \dot n(t) + \langle\phi(t)|\dot\phi(t)\rangle = \sum_{j\in\mathcal{J}}\langle\phi(t)|\hat{\mathrm{B}}^{(\phi_t)}_j|\phi(t)\rangle\text{ ,}\\
 \dot n(t) + \langle\dot\phi(t)|\phi(t)\rangle = \sum_{j\in\mathcal{J}}\langle\phi(t)|\hat{\mathrm{B}}^{(\phi_t)}_j|\phi(t)\rangle\text{ .}
\end{align}
Adding both equations and using $\dot n(t) = \langle\dot\phi(t)|\phi(t)\rangle+\langle\phi(t)|\dot\phi(t)\rangle$ we obtain
\begin{align}
 \frac{\dot n(t)}{n(t)} &= \frac{4}{3}\sum_{j\in\mathcal{J}}\left[\left(\mathbb{E}_j^{(\phi_t,1)}\right)^2-\mathbb{E}_j^{(\phi_t,2)}\right]\\
 &=-\frac{4}{3}\sum_{j\in\mathcal{J}} \text{Var}^{(\phi_t)}_j\text{ ,}\label{NormEvol}
\end{align}
where $\text{Var}^{(\phi_t)}_j$ stands for the variance of the measurement $\hat{\mathrm{O}}_j$ on the state $|\phi(t)\rangle$.
This equation shows that $|\phi(t)\rangle$ is an equilibrium state of Eq. \eqref{ProjLind0} if and only if it has zero variance in all operator measurements and hence
is an eigenstate of all Hermitian operators. To simplify Eq. \eqref{ProjLind0} even further, we write 
\begin{align}
 |\Phi(t)\rangle &\equiv |\phi(t)\rangle \langle\phi(t)|\phi(t)\rangle\text{ ,}\\
 \partial_t|\Phi(t)\rangle &= \sum_{j\in\mathcal{J}}\hat{\mathrm{B}}^{(\Phi_t)}_j|\Phi(t)\rangle\text{ ,}\label{ProjLind1}
\end{align}
and only look at the evolution of the newly defined ``wave function'' $|\Phi(t)\rangle$. The previously determined properties of the equation are unaffected by this change of variables.
Furthermore, as shown in appendix \ref{APPstable}, these fixed point solutions are \textit{asymptotically stable} and small perturbations decay exponentially with $|\vec{a}-\vec{a}'|^2$.

In order to conserve the symmetry of the input state for the eigenstate, the wave function is collapsing to, we can only approximate eigenstates belonging to 
\textit{pairwise distinct} eigenvalues. Therefore, if the symmetry operators commute with the Hermitian operators each distinct eigenstate in the linear decomposition inherits
the symmetry of the input wave function. A proof for the case of fully (anti)-symmetric many-particle wave functions is given in appendix \ref{SymInh}.

%%%%%%%%%%%%%%%%%%%%%%%%%%%%%%%%%%%%%%%%%%%%%%%%%%%%%%%%%%%%%%%%
%%%%%%%%%%%%%%%%%%%%%%%%%%%%%%%%%%%%%%%%%%%%%%%%%%%%%%%%%%%%%%%%
\section{Eigenstate Towing}
\label{eigen_tow}
\begin{figure}[htbp]
\begin{center}
\input{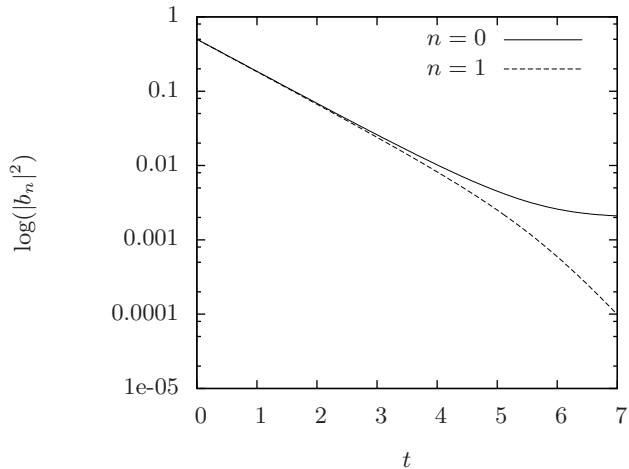}
\caption{Evolution of the probabilities $p_0(t)$ (solid) and $p_1(t)$ (dashed) for $p_0(0) = \frac{501}{1000}$, $p_1(0) = \frac{499}{1000}$ and $p_2(0) = 0$.}
\label{harmex1}
\end{center}
\end{figure}

\begin{figure}[htbp]
\begin{center}
\input{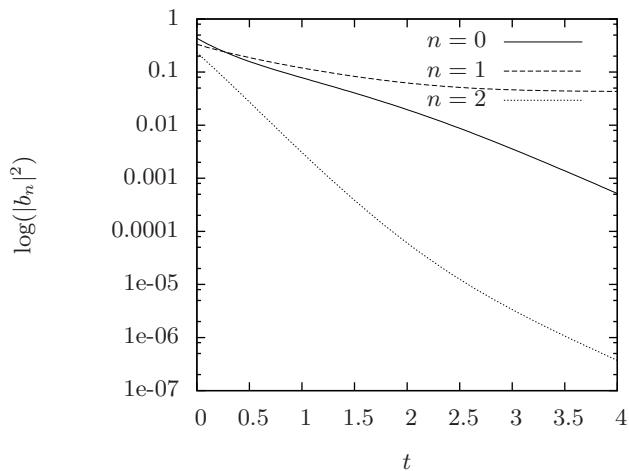}
\caption{Evolution of the probabilities $p_0(t)$ (solid), $p_1(t)$ (dashed) and $p_2(t)$ (dotted) for $p_0(0) = \frac{13}{30}$, $p_1(0) = \frac{10}{30}$ and $p_2(0) = \frac{7}{30}$.}
\label{harmex2}
\end{center}
\end{figure}

\begin{figure}[htbp]
\begin{center}
\input{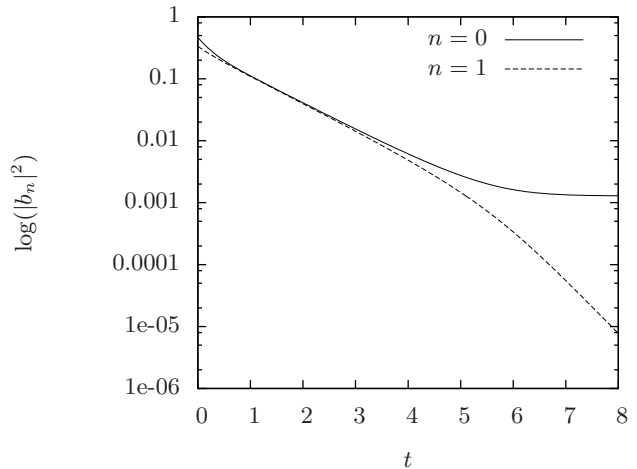}
\caption{Evolution of the probabilities $p_0(t)$ (solid) and $p_1(t)$ (dashed) for $p_0(0) = \frac{14}{30}$, $p_1(0) = \frac{10}{30}$ and $p_2(0) = \frac{6}{30}$.}
\label{harmex3}
\end{center}
\end{figure}

By projecting the dynamics of the density operator on a wave function we have sacrificed the exact compliance with Born's law, 
i.e. in certain cases we may fail to converge to the most probable eigenstate of the decomposition. 
To examplify this issue, let us take the problem of a one-dimensional harmonic oscillator. The only operator we use is the dimensionless Hamilton operator
\begin{align}
 \{\hat{\mathrm{O}}_j\}_{j\in\mathcal{J}} &= \{\hat{\mathrm{H}}\}\text{ ,}\\
 \hat{\mathrm{H}} &= \frac{\hat{v}^2+\hat{x}^2}{2}\text{ ,}
\end{align}
where 
\begin{equation}
 \hat{x} = \frac{\hat{q}}{\sqrt{\frac{\hbar}{m\omega}}}\text{ , } \hat{v} = \frac{\hat{p}}{\sqrt{\hbar m\omega}}\text{ ,}
\end{equation}
are the dimensionless position and momentum operator.
For the representation, we choose the well-known eigenfunctions
\begin{equation}
 \psi_n(x) = \frac{\pi^{-\frac{1}{4}}}{\sqrt{2^n n!}}e^{-x^2/2}H_n(x)
\end{equation}
as a basis such that
\begin{equation}
 \hat{\mathrm{H}}|\psi_n\rangle = (n+1/2)|\psi_n\rangle\text{ .}
\end{equation}
As an initial state we pick 
\begin{equation}
 |\Psi(0)\rangle = b_0 |\psi_0\rangle + b_1 |\psi_1\rangle + b_2 |\psi_2\rangle\text{ ,}
\end{equation}
with $b_0,b_1,b_2\in\mathbb{C}$ and $|b_0|^2+|b_1|^2+|b_2|^2 \equiv p_0+p_1+p_2 = 1$. The dynamics of the coefficients are easily simulated with
Eqs. \eqref{DynSys0} and \eqref{DynSys1}. In Figs. \ref{harmex1}-\ref{harmex3} we show three important examples. In the first example, see Fig. \ref{harmex1},
there are only two nonzero coefficients. One coefficient is choosen slightly bigger than the other, such that the coefficient simulation converges to one
of them. In practice, numerical noise together with a renormalization of the wave function will trigger the collapse to one of the eigenstates.  
In the second example, see Fig. \ref{harmex2}, the three lowest eigenstates participate in the
competition with $p_0>p_1>p_2$. Due to the stronger influence of state $n=0$ on $n=2$ and vice versa the first state $n=1$ survives the competition, although the ground state
is initially the most probable one. In the third example, see Fig. \ref{harmex3}, we show that the system will tend again to the most probable state if a certain bias is exceeded.
In conclusion, the dynamics tend to favor eigenstates in the center of the spectrum of all states $|\psi_{\vec{a}}\rangle$ contained in the linear combination of the input state, i.e.
$\{\vec{a}\in\mathcal{A} \text{ with } |b_{\vec{a}}|^2>0\}$. This is caused by the overall damping rate $\gamma_{\vec{a}}$ for the eigenstate with eigenvalue $\vec{a}$
\begin{equation}
 \gamma_{\vec{a}} = -\sum_{\vec{a}'\in\mathcal{A}}|\vec{a}-\vec{a}'|^2 |b_{\vec{a}'}|^2\text{ .}
 \end{equation}
that is stronger for eigenstates at the boundary than for states in the center of the spectrum.

In order to overcome this difficulty, we now propose the method of \textit{eigenstate towing} (ET), based on the theoretical model described in the previous section, in the following way. 
One starts with a set of operators $\{\hat{\mathrm{O}}^{(0)}_j\}_{j\in\mathcal{J}}$ that have a well-known eigenbasis
$\{|\psi_{\vec{a}_0}\rangle\}_{\vec{a}_0\in\mathcal{A}_0}$ and gradually perturbs the set with the operators $\{\hat{\delta}^{(i)}_j\}_{j\in\mathcal{J}}$ for 
$i\in\{0,1,\dots,M\}\subset\mathbb{N}_0$ to reach the desired problem $\{\hat{\mathrm{O}}_j^{(M)}\}_{j\in\mathcal{J}}$ with corresponding eigenstates 
$\{|\psi_{\vec{a}_M}\rangle\}_{\vec{a}_M\in\mathcal{A}_M}$, 
%or the desired proximity to the phase transition $\kappa\approx \kappa_c$ after $M$ perturbation steps,
i.e.
\begin{equation}
 \hat{\mathrm{O}}_j^{(i)} \equiv \hat{\mathrm{O}}_j^{(0)}+\sum_{k=0}^{i-1} \hat{\delta}_j^{(k)}\text{ .}\label{OpEv}
\end{equation}
It is important to note that the choice of perturbation operators is not limited by
\begin{equation}
 \hat{\delta}^{(i)}_j = \frac{\hat{\mathrm{O}}_j^{(M)}-\hat{\mathrm{O}}_j^{(0)}}{M}\text{ ,}\label{SimChoice}
\end{equation}
but can have any form that fullfills Eq. \eqref{OpEv} for $i=M$. As an example, one can imagine the helium atom in the discrete hyperspherical harmonics' basis such that the matrix representation 
of the Hamilton operator becomes sparse \cite{aquilanti92} and the off-diagonal elements can be treated as perturbations; or one thinks of the quantum phase transition from 
the $U(5)$- to the $SU(3)$-symmetry group in the interacting boson model at $\kappa= \kappa_c$ \cite{feng81}. 
The quantum states, whose progression along the perturbations we would like to follow, are the finite subset of eigenstates 
$\{|\psi_{\vec{a}_0}\rangle\}_{\vec{a}_0\in\mathcal{B}_0}$ with $\mathcal{B}_0\subset\mathcal{A}_0$. One can imagine the procedure as towing the eigenstates along
the line of perturbations. The choice of the subset $\mathcal{B}_0$ is theoretically arbitrary. 
The reason is that from time-independent perturbation theory we know that for infinitesimal perturbations the eigenstates change very little, which implies
that for perturbations of the form \eqref{SimChoice} and in the limit of large $M$ the overlap between the unperturbed $|\psi_{\vec{a}_i}\rangle$ and perturbed eigenstate 
$|\psi_{\vec{a}_{i+1}}\rangle$ satisfies
\begin{equation}
 |\langle\psi_{\vec{a}_i}|\psi_{\vec{a}_{i+1}}\rangle|^2 \lesssim 1\text{ ,}
\end{equation}
with $i\in\{0,1,\dots,M-1\}$. This ensures that the input wave function $|\psi_{\vec{a}_i}\rangle$ will converge to the eigenstate $|\psi_{\vec{a}_{i+1}}\rangle$
using the operators $\{\hat{\mathrm{O}}_j^{(i+1)}\}_{j\in\mathcal{J}}$ for the dynamics in Eq. \eqref{ProjLind1}.
However, depending on the (un-)perturbed operators, excited states are often more affected by perturbations and hence subject to bigger changes. 
Therefore, it may be necessary to adjust the set of perturbation operators to every state which is going to be simulated $\{\hat{\delta}^{(i_{\vec{a}})}_j\}_{j\in\mathcal{J}}$ with 
$i_{\vec{a}}\in\{0,1,\dots,M_{\vec{a}}\}$ and $\vec{a}\in\mathcal{B}$ to establish the reported convergence. 
In practice, the adjustments may only require a larger number $M_{\vec{a}}$ whereas the functional form of the operator stays unchanged, which leads to a higher
computational effort to simulate the behavior of these states. In general, it is difficult to determine $M$, $M_{\vec{a}}$ or the total set 
$\{\hat{\delta}_j^{(i_{\vec{a}})}\}_{j\in\mathcal{J}}$ for a given state and problem in advance. Therefore, 
one should check the convergence to the correct eigenstate by decreasing the perturbation increment and comparing the results to the larger perturbation increment. 
However, since the ET simulation does not require the results from the other computations, each single one can be performed independently and in parallel, which can
represent a significant advantage if the required resources for the calculations are available.
A natural stopping criterion for the convergence in the $i_{\vec{a}}$-th perturbation step of Eq. \eqref{ProjLind1} is the $L^2$-norm of the temporal derivative of the wave function,
whose exponential convergence will be shown for a concrete example in section \ref{ESQPT}.

%%%%%%%%%%%%%%%%%%%%%%%%%%%%%%%%%%%%%%%%%%%%%%%%%%%%%%%%%%%%%%%%
%%%%%%%%%%%%%%%%%%%%%%%%%%%%%%%%%%%%%%%%%%%%%%%%%%%%%%%%%%%%%%%%
\section{Numerical implementation}
\label{num_impl}
With the intention to show a practical application of the proposed algorithm from the previous section we use the (semi)-implicit Crank-Nicolson method for the numerical implementation, which is based on the trapezoidal rule and reads in our case
\begin{equation}
 \frac{\vec{\Phi}_{t+\delta t}-\vec{\Phi}_{t}}{\delta t} = \sum_{j\in\mathcal{J}}\frac{1}{2}\left[\mathrm{B}_j^{(\Phi_{t+\delta t})}\vec{\Phi}_{t+\delta t}+\mathrm{B}_{j}^{(\Phi_t)}\vec{\Phi}_{t}\right]\text{ ,}
\end{equation}
where $\mathrm{B}_j^{(\Phi_t)}$ is the matrix representation of the Hermitian operator $\hat{\mathrm{B}}_j^{(\Phi_t)}$ of Eq. \eqref{B_OP}.
The issue with this procedure is that we do not know the expectation values $\mathbb{E}_j^{(\Phi_{t+\delta t},i)}$ and hence would need to approximate them to first order, 
such that we preserve the second order accuracy, i.e.
\begin{equation}
 \mathbb{E}_j^{(\Phi_{t+\delta t},i)}\approx \mathbb{E}_j^{(\Phi_t,i)} + 2\delta t\sum_{j'\in\mathcal{J}}\vec{\Phi}^T_t\left(\mathrm{O}_j\right)^i\mathrm{B}^{(\Phi_t)}_{j'}\vec{\Phi}_t\text{ .}
\end{equation}
However, practical numerical examples have revealed that a zeroth order approximation 
\begin{equation}
 \mathrm{B}_j^{(\Phi_{t+\delta t})}\approx \mathrm{B}_j^{(\Phi_t)}\text{ ,}
\end{equation}
is more stable, such that a significantly larger time-step can be used and a faster convergence is obtained. 
To solve the system of algebraic equations we use a Cholesky decomposition of our real symmetric, positive-definite matrix 
\begin{equation}
 \mathrm{M}^{(\Phi_{t+\delta t})}\equiv \mathbb{I}+\frac{\delta t}{2}\sum_{j\in\mathcal{J}}\mathrm{B}^{(\Phi_t)}_j\text{ .}
\end{equation}
The matrix is positive-definite, since for any eigenvector $\vec{\Psi}_{\vec{a}}$ of the set of matrices $\{\mathrm{O}_j\}_{j\in\mathcal{J}}$ we know 
\begin{align}
 \mathrm{B}_j^{(\Psi_{\vec{a}})}\Psi_{\vec{a}} &= \vec{0}\text{ ,}\\
 \mathrm{M}^{(\Psi_{\vec{a}})}\vec{\Psi}_{\vec{a}} &= \vec{\Psi}_{\vec{a}}\text{ ,}
\end{align}
i.e. $M^{(\Psi)}$ only has the positive eigenvalue $+1$. 
For a sparse matrix representation of the observables, the Crank-Nicolson method offers the advantage of higher stability, and hence larger time-steps compared to an explicit method, 
whereas the computational costs for each step are compatible due to the Cholesky decomposition. In section \ref{benchmarking} we further analyze the convergence and scaling properties 
of our algorithm compared to the MRRR method \cite{dhillon97} for eigenvector computations of tridiagonal matrices, which arise in the example of an excited-state quantum phase transition 
that will be discussed in the next section.

%%%%%%%%%%%%%%%%%%%%%%%%%%%%%%%%%%%%%%%%%%%%%%%%%%%%%%%%%%%%%%%%
%%%%%%%%%%%%%%%%%%%%%%%%%%%%%%%%%%%%%%%%%%%%%%%%%%%%%%%%%%%%%%%%
\section{Application to Quantum Phase Transitions}
\label{ESQPT}
\begin{figure}[htbp]
\begin{center}
\input{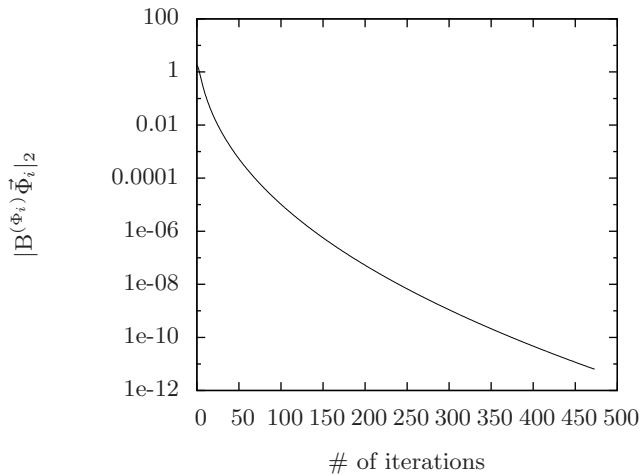}
\caption{Evolution of the stopping criterion for the collapse algorithm using $\delta t = 1.1$, $N=80$ molecules targeting the $8$-th eigenvector for $c=j$, $\kappa=0.1$, $\omega_0 = 1$ and $\omega = 2$.}
\label{RMSConv}
\end{center}
\end{figure}

\begin{figure}[htbp]
\begin{center}
\input{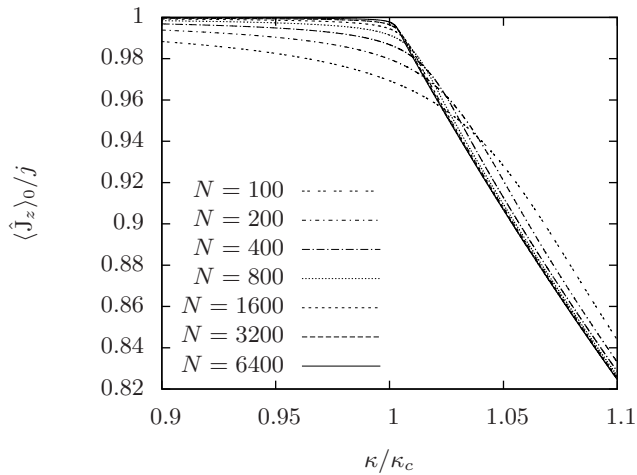}
\caption{Scaled atomic inversion as a function of the interaction strength for the ground state eigenvector using different numbers of molecules $N$, $c=j$, $\omega_0 = 1$ and $\omega = 2$.}
\label{OP}
\end{center}
\end{figure}

\begin{figure}[htbp]
\begin{center}
\input{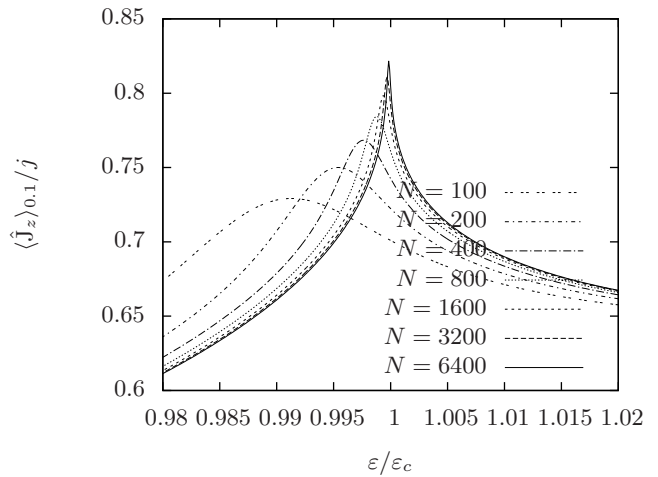}
\caption{Atomic inversion as a function of the scaled energy for the spectrum ratio $q=0.1$ using different numbers of molecules $N$, $c=j$, $\omega_0 = 1$ and $\omega = 2$.}
\label{AI_k=1}
\end{center}
\end{figure}

\begin{figure}[htbp]
\begin{center}
\input{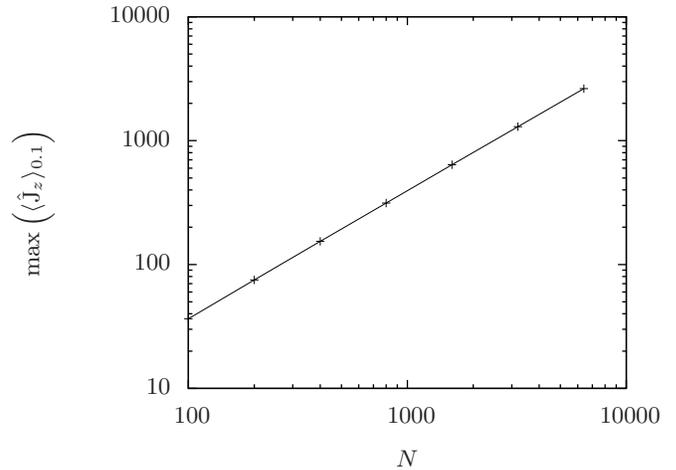}
\caption{Maximum atomic inversion as a function of the number of molecules for the spectrum ratio $q=0.1$, $c=j$, $\omega_0 = 1$ and $\omega = 2$, measuring a slope $\beta_q/\nu_q \approx 1.03$.}
\label{critexpEx}
\end{center}
\end{figure}

\begin{figure}[htbp]
\begin{center}
\input{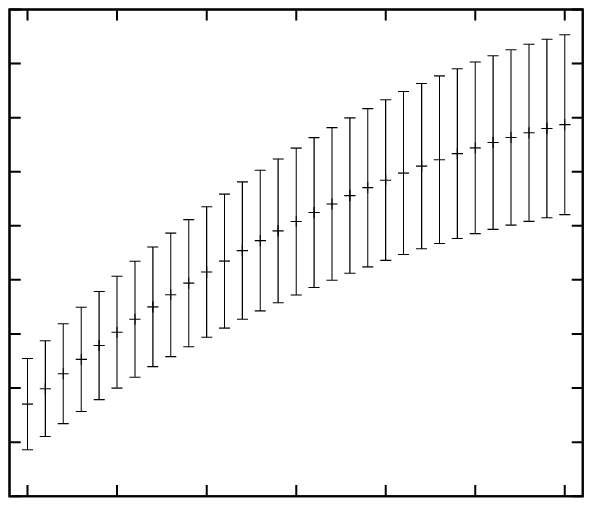}
\caption{Change in the critical exponent ratio $\beta_q/\nu_q$ of the atomic inversion as function of the spectrum ratio $q=k/N$ with $c=j$, $\omega_0 = 1$ and $\omega = 2$, and
errors showing the $95\%$ confidence interval of the linear regression model.}
\label{critexp}
\end{center}
\end{figure}

As a practical example we consider the excited-state quantum phase transition (see Refs. \cite{caprio08,cejnar11}) in the Jaynes-Cummings model, further explained in Refs. 
\cite{jaynes63,tavis68,arias11}.
The quantum optical model depicts the behavior of $N$ identical two-level molecules coupled to a single-mode radiation field. The model is simple but still covers the main
features of the class of problems we are interested in. The Hamilton operator can be written as
\begin{equation}
 \hat{\mathrm{H}}_{\kappa} = \omega_0\hat{\mathrm{J}}_z + \omega\hat{\mathrm{b}}^{\dagger}\hat{\mathrm{b}} + \frac{\kappa}{\sqrt{4j}}\left[\hat{\mathrm{b}}\hat{\mathrm{J}}_+ + \hat{\mathrm{b}}^{\dagger}\hat{\mathrm{J}}_-\right]\text{ ,}
\end{equation}
where $2j$ equals the number of molecules $N$ and $\kappa$ determines the interaction strength between
the radiation field and the molecules. The operators fullfill the usual bosonic and $SU(2)$ commutation relations
\begin{align}
 \left[\hat{\mathrm{J}}_z,\hat{\mathrm{J}}_{\pm}\right]&=\pm\hat{\mathrm{J}}_{\pm}\text{ ,}\\
 \left[\hat{\mathrm{J}}_+,\hat{\mathrm{J}}_-\right]&=2\hat{\mathrm{J}}_z\text{ ,}\\
 \left[\hat{\mathrm{b}},\hat{\mathrm{b}}^{\dagger}\right]&=1\text{ .}
\end{align}
As a basis we choose the states $|n\rangle|j,m\rangle$ which are eigenstates of the non-interacting system, such that
\begin{equation}
 \hat{\mathrm{H}}_0\left(|n\rangle|j,m\rangle\right) = (\omega_0 m+\omega n)\left(|n\rangle|j,m\rangle\right)\text{ ,}
\end{equation}
and at the same time eigenstates of the operator $\hat{\mathrm{J}}^2$ with 
\begin{align}
 \hat{\mathrm{J}}^2 &\equiv \hat{\mathrm{J}}_z^2+(\hat{\mathrm{J}}_+ \hat{\mathrm{J}}_- + \hat{\mathrm{J}}_- \hat{\mathrm{J}}_+)/2\text{ ,}\\
 \hat{\mathrm{J}}^2\left(|n\rangle|j,m\rangle\right) &= j(j+1)\left(|n\rangle|j,m\rangle\right)\text{ ,}
\end{align}
in analogy to the angular momentum formalism.
Another important observation is that both $\hat{\mathrm{b}}^\dagger\hat{\mathrm{b}}+\hat{\mathrm{J}}_z$ and $\hat{\mathrm{J}}^2$ commute with $\hat{\mathrm{H}}_{\kappa}$. 
Hence we choose the energy eigenstates $|j,c,\epsilon\rangle$ to be eigenstates of these two operators as well, i.e.
\begin{align}
\hat{\mathrm{J}}^2 |j,c,\epsilon\rangle &= j(j+1)|j,c,\epsilon\rangle\text{ ,}\\
\left(\hat{\mathrm{b}}^\dagger\hat{\mathrm{b}}+\hat{\mathrm{J}}_z\right)|j,c,\epsilon\rangle &= c|j,c,\epsilon\rangle\text{ ,}\\
\hat{\mathrm{H}}_{\kappa}|j,c,\epsilon\rangle &= \epsilon|j,c,\epsilon\rangle\text{ .}
\end{align}
Therefore we write the eigenstate as
\begin{equation}
  |j,c,\epsilon\rangle = \sum_{n=c-j}^{c+j} A_n^{(j,c,\epsilon)} |n\rangle|j,c-n\rangle\text{ ,}
\end{equation}
where the sum can only run along $n\in\mathbb{N}_0$ number of bosons. The eigenvector problem hence transforms into determining the coefficient vector $A_n^{(j,c,\epsilon)}$, and
the matrix representation of $\hat{\mathrm{H}}_{\kappa}$ becomes tridiagonal in this basis. The reason is that the interaction term can only alter the parameter $n$ by $\pm 1$, which leads
to a system of equations 
\begin{align}
 &\frac{\kappa}{\sqrt{4j}}\left[\sqrt{n}C_{j,c-n}A_{n-1}^{(j,c,\epsilon)}+\sqrt{n+1}C_{j,c-n-1}A_{n+1}^{(j,c,\epsilon)}\right]\notag\\
 &+\left[(c-n)\omega_0+n\omega-\epsilon\right]A_n^{(j,c,\epsilon)} = 0\text{ ,}
\end{align}
where $C_{j,c-n} \equiv \sqrt{j(j+1)-(c-n)(c-n+1)}$.
In summary, this means that we choose
\begin{align}
 \hat{\mathrm{O}}^{l_k} &\equiv \op{H}_{\frac{l}{M_k}\kappa}\\
 l &\in \{1,2,\dots,M_k\}\text{ ,}\\
 \{|\psi\rangle_{a_0}\}_{a_0\in\mathcal{A}_0} &\equiv \{|n\rangle|j,c-n\rangle\}_{n\in\{c-j,\dots,c+j\}}\text{ ,}\\
 \op{O}_{(0)}|n\rangle|j,c-n\rangle &= \left[(c-n)\omega_0+n\omega\right]|n\rangle|j,c-n\rangle\text{ ,}\\ 
 \op{\delta}^{(l_k)} &\equiv \frac{l}{M_k}\frac{\kappa}{\sqrt{4j}}\left[\hat{\mathrm{b}}\hat{\mathrm{J}}_+ + \hat{\mathrm{b}}^{\dagger}\hat{\mathrm{J}}_-\right]\text{ ,}\\
 k &\in \{c-j,c-j+1,\dots,c+j\}\text{ ,}
\end{align}
For the numerical implementation this choice of basis functions implies that we pick an initial vector
$\vec{\Phi}_0$ from the set of standard basis vectors $\{\hat{e}_0,\hat{e}_2,\dots,\hat{e}_{2j}\}$
which span the space $\mathbb{R}^{2j+1}$, for instance $\vec{\Phi}_0 \equiv \vec{\Phi}_{k,0} = \hat{e}_k$, and apply the first ($l=1$) measurement dynamical evolution, 
defined in section \ref{deriv}, with the matrix
\begin{align}
  \left(\mathrm{H}_{\frac{1}{M_k}\kappa}\right)_{ii'} &= \left[(j-i)\omega_0+(c-j+i)\omega\right]\delta_{i,i'}\notag\\
  &+\frac{1}{M_k}\frac{\kappa}{\sqrt{4j}}\left[\sqrt{c-j+i}C_{j,j-i}\delta_{i,i'-1}\right.\notag\\
  &\left. +\sqrt{c-j+i+1}C_{j,j-i-1}\delta_{i,i'+1}\right]\text{ .}
\end{align}
The matrix representation of the time evolution operator, defined in Eq. \eqref{B_OP}, that is applied in the first operation of the algorithm looks like
\begin{align}
 B^{(\phi_{k,0})} &= 2 \mathbb{E}^{(\phi_{k,0},1)} \mathrm{H}_{\frac{1}{M_k}\kappa}-\mathrm{H}_{\frac{1}{M_k}\kappa}^2-\mathbb{E}^{(\phi_{k,0},2)}\mathbb{I}\text{ ,}\\
 \mathbb{E}^{(\phi_{k,0},1)} &= \left(\hat{e}_k\right)^T \mathrm{H}_{\frac{1}{M_k}\kappa} \hat{e}_k = (j-k)\omega_0+(c-j+k)\omega\text{ ,}\\
 \mathbb{E}^{(\phi_{k,0},2)} &= \left(\hat{e}_k\right)^T \mathrm{H}^2_{\frac{1}{M_k}\kappa} \hat{e}_k = \sum_{i=0}^{2j} \left(\mathrm{H}_{\frac{1}{M_k}\kappa}\right)_{ki}\left(\mathrm{H}_{\frac{1}{M_k}\kappa}\right)_{ik}\text{ .}
\end{align}
The only part that is going to change when making the next time-step of the algorithm are the expectations values, which hence changes the matrix $B$, which is used for the next 
time-step. This procedure will be iterated until the desired convergence, based on the criterion exemplified in Fig. \ref{RMSConv}, is reached. 
The resulting vector $\vec{\Phi}_{t_{\text{conv}}}$ is then called $\vec{\Phi}_0$ again and the next measurement dynamics ($l=2$) with the matrix 
$\mathrm{H}_{\frac{2}{M_k}\kappa}$ is started. This process is performed until $l=M_k$.
In the case of an excited-state as opposed to a usual quantum phase transition one observes the transition and hence the critical scaling behavior with respect to an increase in the 
interaction strength $\kappa$ not only in the ground state but also for eigenstates belonging to larger energy eigenvalues, cf. Ref. \cite{caprio08}. 
Therefore, we analyze the scaling behavior of the atomic inversion $\langle\hat{\mathrm{J}}_z\rangle$ not just as a function of the number of identical two-level molecules $N=2j$ 
in the ground state, where it serves as an order parameter, but also as a function of the spectrum ratio $q\equiv k/N$. $k$ is the level of the excited-state eigenvector $\vec{\Phi}_k$
in the vectorspace $\mathbb{R}^{N+1}$. The scaled atomic inversion,
\begin{equation}
 \langle \hat{\mathrm{J}}_z\rangle_q/j =  1-\frac{1}{j}\sum_{i=0}^N \Phi_{k,i}^2 i\text{ ,}
\end{equation}
shown in Fig. \ref{OP}, decreases from one above the theoretical transition point $\kappa_c=\sqrt{\frac{(\omega-\omega_0)^2}{2}}$ \cite{arias11} for the ground state ($q=0$). 
In addition, we observe the finite size effects for different numbers of molecules.
In the case of an excited state, the atomic inversion also shows critical scaling as a function of the interaction strength or the scaled energy $\varepsilon\equiv \epsilon/j$, as demonstrated in
Ref. \cite{arias11}. Although it does not serve as an order parameter to distinguish the two phases, see Fig. \ref{AI_k=1}, one can for instance still analyze its finite size scaling
\begin{equation}
 \langle \hat{\mathrm{J}}_z\rangle_{q} = N^{\beta_q/\nu_q}\mathcal{F}_{J_z,q}\left[(\varepsilon-\varepsilon_c)N^{1/\nu_q}\right] \text{ ,}
\end{equation}
where $\varepsilon_c$ is the critical scaled energy. To measure the critical exponent $\beta_q/\nu_q$ as a function of the spectrum ratio $q\equiv k/N$,
we measure the slope of the linear model fit of the maximum atomic inversion plotted against the number of molecules, as shown in Fig. \ref{critexpEx}. The results are shown in Fig. 
\ref{critexp} and demonstrate a general trend for an increase of $\beta_q/\nu_q$ with an increasing spectrum ratio $q$, which means that the critical exponents are not independent
of the spectrum ratio. The errors depict the $95\%$ confidence interval for the slope parameter of the fitted 
linear regression model. Note that the error bars and the small relative change of $2.5\%$ may suggest that the critical exponent ratio would be universal. 
However, we have refocused our simulation efforts around the critical points $\kappa_c(q)$ to reduce the main source of error, which is to underestimate the maximum value due
to the discrete scanning of $\kappa$ values, and the general trend as well as the measured slopes are unchanged. Furthermore, we changed $\omega$ and found that, as expected, the
critical exponent ratios do not change.

\section{Benchmarking}
\label{benchmarking}
\begin{figure}[htbp]
\begin{center}
\input{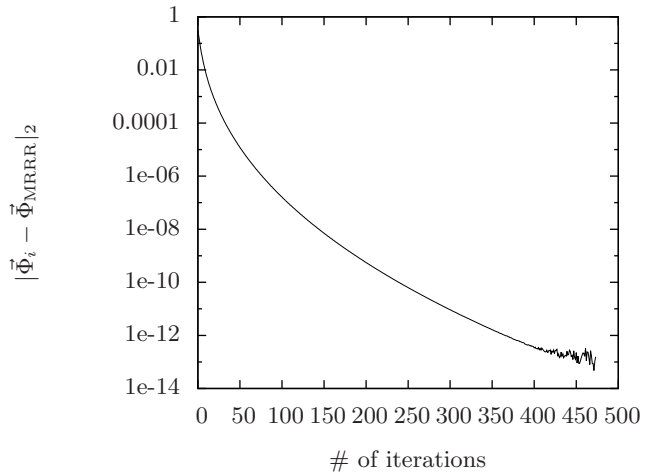}
\caption{Convergence of ET algorithm using $\delta t = 1.1$ for $N=80$ molecules towards the $8$-th eigenvector calculated by LAPACK MRRR algorithm \cite{dhillon97,dhillon06} for $c=j$, $\kappa=0.1$, $\omega_0 = 1$ and $\omega = 2$.}
\label{EVecConv}
\end{center}
\end{figure}

\begin{figure}[htbp]
\begin{center}
\input{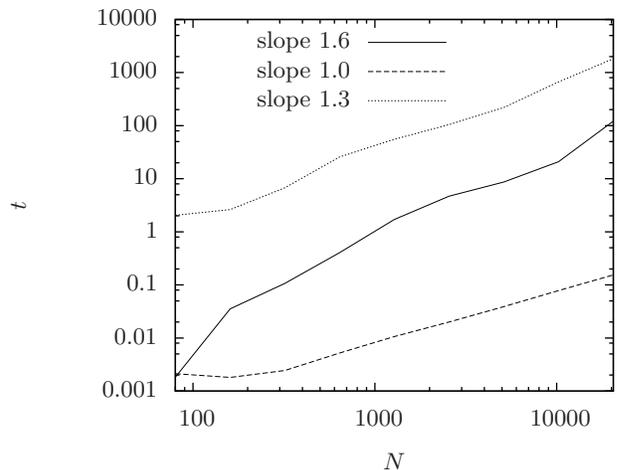}
\caption{Scaling behavior of CPU time for LAPACK MRRR algorithm \cite{dhillon97,dhillon06} targeting only the $N/10$-th eigenvector (dashed), all eigenvectors (solid) and first-order ET algorithm targeting only the $N/10$-th eigenvector using $\delta t = 1.1$ (dotted) with $c=j$, $\kappa=0.1$, $\omega_0=1$ and $\omega=2$ on $6$ Intel(R) Xeon(R) CPUs E5-1650 v3 3.50GHz with 15.36 MB cache and $12$ processors.}
\label{scal_comp}
\end{center}
\end{figure}

\begin{figure}[htbp]
\begin{center}
\input{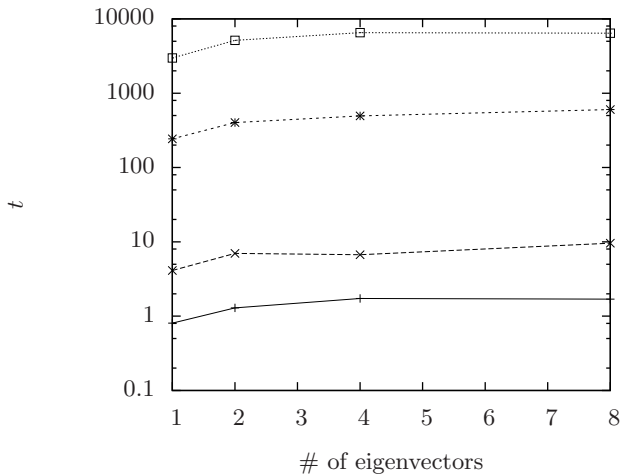}
\caption{Parallel first-order ET algorithm targeting $n$ eigenvectors ($\{N/10,N/10-1,\dots,N/10-(n-1)\}$) with $N=80$ (solid), $640$ (long dashed), $5120$ (short dashed), $20480$ (dotted) using $\delta t = 1.1$ with $c=j$, $\kappa=0.1$, $\omega_0=1$, $\omega=2$ and the OpenMP maximally parallelized PARDISO \cite{schenk04,schenk06,karypis98} for the implicit time-step on $6$ Intel(R) Xeon(R) CPUs E5-1650 v3 3.50GHz with 15.36 MB cache and $12$ processors.}
\label{par_scal_omp}
\end{center}
\end{figure}

In order to compare our method to other state-of-the-art eigensolvers on the problem of excited-state quantum phase transitions in the Jaynes-Cummings model, described in the previous section, 
which essentially boils down to the diagonalization of a tridiagonal matrix, 
we choose one of the fastest, but still accurate algorithms, the MRRR algorithm \cite{dhillon97,dhillon06}, in the publicly available 
\textit{Linear Algebra PACKage} (LAPACK) via the \textit{INTEL Math Kernel Library} version 11.0.3. \cite{demmel08}. In Fig. \ref{EVecConv} we show the convergence of a 
higher-order eigenvector to the result computed by the MRRR algorithm, implemented in the LAPACK library \cite{dhillon06}. First of all,
one observes that the convergence is not exponential throughout the whole dynamics. Nevertheless, from Eq. \eqref{exp_dec} we infer that the exponential convergence only holds close to the 
exact result, which can be confirmed by looking at the graph between $300$ and $400$ iterations. 
In addition, the renormalization which is performed every time-step affects the dynamics and hence also the convergence properties, 
especially in the beginning where the variance and consequently the change in the norm of the vector, see Eq. \eqref{NormEvol}, are quite large.
Secondly, due to the accumulation of numerical noise, we find that the eigenvectors do not converge with machine precision to the MRRR result, but instead observe a precision loss with an increased number of entries in the matrix representation. The reason is 
the higher number of floating point operations in the matrix-vector multiplications. This also causes the tremor at the end of the line in Fig. \ref{EVecConv}. 
As mentioned at the end of section \ref{eigen_tow}, Fig. \ref{RMSConv} shows the evolution of the approximated temporal derivative that is used as a stopping criterion for the algorithm.
The behavior is similar to the convergence of the eigenvector in the previous figure. As before, one observes the exponential decay at the end of the evolution, 
between $300$ and $450$ iterations.
In Fig. \ref{scal_comp} we compare the scaling behavior of our ET algorithm with the MRRR algorithm that
computes all and only one specific eigenvector. Using our method to sequentially calculate all eigenvectors and -values mainly means that the computation times are multiplied by $N$, 
the number of eigenvectors in the problem, which is why we did not include it in this figure. 
In general, we find the expected linear scaling behavior for the single-eigenvector LAPACK MRRR algorithm. For the LAPACK algorithm 
that computes all eigenvectors, we found a slope of $1.6$ although in theory we would expect $2.0$. We attribute the difference to parallelization, which has probably affected 
the computation times more effectively for larger matrix sizes. For our algorithm we observe a scaling with a slope of $1.3$ that lies in the middle between the LAPACK algorithms. 
The scaling behavior of the ET algorithm is mainly caused by the increased number of perturbation steps, which are required to ensure
convergence to the correct eigenvector in the spectrum. In addition, we demonstrate the parallelizability of the ET algorithm in Fig. \ref{par_scal_omp} for different matrix sizes and different
numbers of computed excited-state eigenvectors. The reason the curves for a given $N$ are not perfectly horizontal is the OpenMP maximally-parallelized ``parallel direct solver'' (PARDISO) 
\cite{schenk04,schenk06,karypis98}, which slows down if less processors per eigenvector computation are available. We have tested other solving modes of PARDISO, but did not find
another scaling behavior.

%%%%%%%%%%%%%%%%%%%%%%%%%%%%%%%%%%%%%%%%%%%%%%%%%%%%%%%%%%%%%%%%
%%%%%%%%%%%%%%%%%%%%%%%%%%%%%%%%%%%%%%%%%%%%%%%%%%%%%%%%%%%%%%%%
\section{Conclusion}
\label{concl}
Starting from a physically motivated perspective on the problem of finding eigenstates of a set of Hermitian operator with discrete spectra, 
we developed a nonlinear Schr\"{o}dinger equation for an isolated quantum mechanical system by projecting the decoherent collapse dynamics 
of the Lindblad equation onto a wave function. The measurements favor the collapse to eigenstates based on Born's law instead of the hierarchy of eigenvalues,
which is the case in most of the other eigensolvers \cite{lanczos50,bathe76,hammond94,feng96,lehto07}. The method, discussed in Ref. \cite{tackett02}, gives the possibility
to target eigenstates with other (arbitrary) selection criteria, which would include Born's law. However, it has only cubic convergence, whereas our nonlinear Schr\"{o}dinger
equation offers exponential convergence to the eigenstate. We showed that the unique stable equilibria of the obtained equation are 
given by the eigenstates of the observables and discussed the eigenstate towing as an application to approximate these states. Thereby, one makes use of the time-independent
perturbation theory to follow the progression of an arbitrary subset of eigenstates along a line of perturbations. On the one hand, it gives the possibility to see the effect of 
stronger interactions on excited eigenstates or states with certain localization properties, such as edge or surface states, 
which is important for excited-state quantum phase transitions \cite{caprio08} and was demonstrated in paragraph \ref{ESQPT} for the Jaynes-Cummings model system 
\cite{jaynes63,tavis68,arias11}. On the other hand, one is able to track the progression of each eigenstate individually and in parallel without the exchange of information between the different computations, 
which allows an efficient calculation on multi-processor clusters, as seen in paragraph \ref{benchmarking} using the (semi-)implicit Crank-Nicolson method as a specific numerical implementation,
discussed in section \ref{num_impl}.
Similar to other methods, such as Ref. \cite{hammond94}, a simple implementation of symmetries, either directly in the input state or explicitly through the usage of the corresponding symmetry operator in the dynamics,
is feasible.
This can help to further increase the convergence rate, since the distance between the eigenvalues of states contained in the decomposition of the input state may increase or is enhanced
by the explicit symmetry operator usage.

As future work, we want to examine other decoherent evolution equations with stronger coupling between the quantum system and the measurement apparatus as well as 
stochastic numerical simulation techniques to make higher-dimensional computations feasible. 

\begin{acknowledgments}
Financial support from the European Research Council (ERC) Advanced Grant 319968-FlowCCS is kindly acknowledged.
In addition, we would like to thank J.-D. Debus for helpful discussions.
\end{acknowledgments}

\bibliography{lit}{}

\appendix

\section{Stability of equilibria}
\label{APPstable}
Without loss of generality we write $|\Phi(t)\rangle = \sum_{\vec{a}\in\mathcal{A}} b_{\vec{a}}(t)|\psi_{\vec{a}}\rangle$ and shift the dynamics into the coefficients $b_{\vec{a}}$. 
Based on Eq. \eqref{ProjLind1} we obtain the following system of coupled non-linear, first-order differential equations for the coefficients
\begin{equation}
 \frac{\dot b_{\vec{a}}(t)}{b_{\vec{a}}(t)} = -\frac{\sum\limits_{\vec{a}'\in\mathcal{A}} |b_{\vec{a}'}(t)|^2 |\vec{a}'-\vec{a}|^2}{\sum\limits_{\vec{a}'\in\mathcal{A}} |b_{\vec{a}'}(t)|^2}\text{ .}\label{CEQS}
\end{equation}
The \textit{unique asymptotically stable equilibria} for this system of differential equations are given by
\begin{equation}
 b_{\vec{a}}(t) = \delta_{\vec{a},\vec{a}'}b_{\vec{a}'}(0)\text{ .}\label{EquilSol}
\end{equation}

To proof this statement, we transform the system by using the definition 
$b_{\vec{a}}(t) \equiv x_{\vec{a}}(t)+\imath y_{\vec{a}}(t)$ with $\left(x_{\vec{a}}(t),y_{\vec{a}}(t)\right)\in \mathcal{B}^2_1\subset \mathbb{R}^2$ 
(assuming the initial state $|\Phi\rangle$ is normalized) into an even bigger, but real system of differential equations
\begin{align}
 \frac{\dot x_{\vec{a}}}{x_{\vec{a}}} &= -\frac{\sum\limits_{\vec{a}'\in\mathcal{A}} [x_{\vec{a}'}^2(t)+y_{\vec{a}'}(t)^2] |\vec{a}'-\vec{a}|^2}{\sum\limits_{\vec{a}'\in\mathcal{A}} [x_{\vec{a}'}^2(t)+y_{\vec{a}'}(t)^2]}\text{ ,}\label{DynSys0}\\
 \frac{\dot y_{\vec{a}}}{y_{\vec{a}}} &= -\frac{\sum\limits_{\vec{a}'\in\mathcal{A}} [x_{\vec{a}'}^2(t)+y_{\vec{a}'}(t)^2] |\vec{a}'-\vec{a}|^2}{\sum\limits_{\vec{a}'\in\mathcal{A}} [x_{\vec{a}'}^2(t)+y_{\vec{a}'}(t)^2]}\text{ .}\label{DynSys1}
\end{align}
To analyze the system we define the solution vector $\vec{u}(t)$ which contains the real and imaginary function of each coefficient, i.e. it has $2\times |\mathcal{A}|$ entries. 
For simplicity, we label the eigenvalues by integers, i.e. $\mathcal{A}=\{\vec{a}^{(0)},\vec{a}^{(1)},\vec{a}^{(2)},\dots\}$, such that we can write
\begin{align}
 \vec{u}(t) &= \{x_{\vec{a}^{(0)}}(t),y_{\vec{a}^{(0)}}(t),x_{\vec{a}^{(1)}}(t),y_{\vec{a}^{(1)}}(t),\dots\}\text{ ,}\\
 &\equiv \{x_0(t),y_0(t),x_1(t),y_1(t),x_2(t),y_2(t),\dots\}\text{ .}
\end{align}
The dynamical system, Eqs. \eqref{DynSys0} and \eqref{DynSys1}, can hence be summarized as
\begin{equation}
 \dot{\vec{u}}(t) = \vec{F}(\vec{u})\text{ .}
\end{equation}
It is relatively easy to see that \textit{if and only if} $\vec{u}^*$ is the equilibrium solution from Eq. \eqref{EquilSol} the right hand side $\vec{F}(\vec{u}^*)$ will vanish. 

Let us assume without loss of generality that in Eq. \eqref{EquilSol} $\vec{a}'=\vec{a}^{(0)}$ then $\vec{u}^* = \{\Re[b_{\vec{a}^{(0)}}],\Im[b_{\vec{a}^{(0)}}],0,\dots,0\}$ 
and the only functions one needs to analyze are $F_0$ and $F_1$, 
since apparently $F_i = 0$ for $i\geq 2$. $F_0$ and $F_1$ are also zero, since they only contain summands from functions $u_i$ with $i\geq 2$ which are zero. 

To determine the stability properties of the equilibrium $\vec{u}^*$, we linearize the problem with respect to small perturbations around the equilibrium solution and 
look at the Jacobian of the vector field $\vec{F}$ at the point $\vec{u}^*$, as described in Ref. \cite{wiggins03}. The resulting matrix is diagonal and its entries are
\begin{align}
-\text{Diag}\left[J_{\vec{F}}(\vec{u}^*)\right] &= \{0,0,|\vec{a}^{(1)}-\vec{a}^{(0)}|^2,|\vec{a}^{(1)}-\vec{a}^{(0)}|^2,\notag\\
&|\vec{a}^{(2)}-\vec{a}^{(0)}|^2,|\vec{a}^{(2)}-\vec{a}^{(0)}|^2,\dots\}\text{ .}
\end{align}
Consequently, we have shown the stability along the directions $\vec{a}^{(i)}$, $i\geq 1$ and potential instability along the direction $\vec{a}^{(0)}$. 
However, due to Eq. \eqref{NormEvol} we know that the overall norm has to decrease during the evolution. Consequently, starting with the initial condition 
$\sum_{\vec{a}\in\mathcal{A}}|b_{\vec{a}}(0)|^2 = 1$, we will necessarily
tend towards a final state with $\lim\limits_{t\rightarrow\infty}|b_{\vec{a}}(t)|^2\leq 1$ $\forall \vec{a}\in\mathcal{A}$, which must also hold for the equilibrium solutions.
The small perturbation $\delta\vec{u}$ from the fixed point solution $\vec{u}^*$ decays as
\begin{equation}
 \delta\vec{u}(t) = e^{J_{\vec{F}}(\vec{u}^*)t}\delta\vec{u}(0)\text{ ,}\label{exp_dec}
\end{equation}
which confirms the exponential decay rate proportional to $|\vec{a}^{(0)}-\vec{a}^{(i)}|^2$ with $i\neq 0$.

\section{Symmetry inheritance}
\label{SymInh}
We are interested in the distinct eigenstates $\{|\psi_n\rangle\}_{n\in\mathbb{N}_0}$ of a quantum many-particle system of $N$ indistinguishable particles, such as fermions or bosons, whose dynamics is governed 
by the Hamilton operator $\hat{\mathrm{H}}$. With distinct we mean that these states belong to different eigenvalues $\epsilon_n$ of $\hat{\mathrm{H}}$, i.e.
\begin{equation}
 \hat{\mathrm{H}}|\psi_n\rangle = \epsilon_n |\psi_n\rangle\text{ .}\label{SEEq}
\end{equation}
They \textit{might not} span the whole Hilbert space, i.e. we may have a degeneracy in our system. However, we assume that we can construct a complete basis with some 
orthoganilization procedure on each subspace $\mathcal{S}_n$ corresponding to a specific eigenvalue. Hence, we can write for all $|\phi\rangle\in\mathcal{H}$
\begin{equation}
 |\phi\rangle = \sum_{n\in\mathbb{N}_0} \left(\SumInt_{k\in\mathcal{S}_n} b_{n,k}|\psi_{n,k}\rangle\right) \equiv \sum_n a_n |\psi_n\rangle\text{ ,}\label{SDES}
\end{equation}
where we have used the superposition principle in the definition.

In the case of bosons the eigenstates, which we desire, are fully symmetric $|\psi_n\rangle_+$ whereas for fermions they are fully anti-symmetric $|\psi_n\rangle_-$.
The basis of pairwise permutations of particles $\mathcal{P}$ with $|\mathcal{P}|=\frac{n(n-1)}{2}$ elements spans the whole set of permutations and hence these (anti)-symmtric states need to fullfill
\begin{equation}
\langle \vec{x}_1,\dots,\vec{x}_N|\hat{P}|\psi_n\rangle_{\pm} = (\pm 1)\langle \vec{x}_1,\dots,\vec{x}_N|\psi_n\rangle_{\pm}\text{ ,} 
\end{equation}
for each element $\hat{P}$ in the set $\mathcal{P}$. We make the important assumption that 
\begin{equation}
 [\hat{P},\hat{\mathrm{H}}] = 0\label{PHC}
\end{equation}
holds for all $\hat{P}\in\mathcal{P}$.

We choose the initial state $|\phi\rangle$ to be fully (anti)-symmetric, i.e. $|\phi\rangle = |\phi\rangle_{\pm}$ and
\begin{equation}
 \hat{P}|\phi\rangle_{\pm} = \pm |\phi\rangle_{\pm} \text{ .}
\end{equation}
Hence, together with Eq. \eqref{PHC} we can conclude
\begin{equation}
 \hat{P}\hat{\mathrm{H}}^l |\phi\rangle_{\pm} = \hat{\mathrm{H}}^l \hat{P}|\phi\rangle_{\pm} = \pm \hat{\mathrm{H}}^l |\phi\rangle_{\pm} \text{ ,}
\end{equation}
for all $l\in\mathbb{N}_0$.
From Eq. \eqref{SEEq} we infer that $\hat{P}|\psi_n\rangle \equiv |\psi_n^P\rangle$ is again an eigenstate to eigenvalue $\epsilon_n$.
With formula \eqref{SDES} and the previous equations we can set up a system of linear equations for all $\hat{P}\in\mathcal{P}$, which reads
\begin{align}
 \sum_{n\in\mathbb{N}_0}  \tilde{a}_n^P \epsilon_n^l = 0 \text{ } \left(\forall l\in\mathbb{N}_0\right)\text{ ,}\\
 \tilde{a}_n^P \equiv a_n (|\psi_n\rangle \mp |\psi_n^P\rangle)\text{ .}
\end{align}
Trying to solve this system for the coefficient vector $(\tilde{a}_0^P,\tilde{a}_1^P,\dots)^T$, we write it as matrix-vector product
\begin{equation}
 \begin{pmatrix}
1 & 1 & 1 & 1 & 1 & \dots \\
\epsilon_0 & \epsilon_1 & \epsilon_2 & \epsilon_3 & \epsilon_4 & \dots \\
\epsilon_0^2 & \epsilon_1^2 & \epsilon_2^2 & \epsilon_3^2 & \epsilon_4^2 & \dots \\
\epsilon_0^3 & \epsilon_1^3 & \epsilon_2^3 & \epsilon_3^3 & \epsilon_4^3 & \dots \\
\vdots & \vdots & \vdots & \vdots & \vdots & \dots
\end{pmatrix}
 \begin{pmatrix}
\tilde{a}_0^P \\
\tilde{a}_1^P \\
\tilde{a}_2^P \\
\tilde{a}_3^P \\
\vdots 
\end{pmatrix}
= 0 \label{VMEQS}
\end{equation}
and realize that our matrix $M$ is a Vandermonde matrix \cite{horn91}. Therefore, we can easily write a formula for the determinant of $M$, given by
\begin{equation}
 \text{Det} M = \prod_{0\leq i < j \leq \infty} (\epsilon_i-\epsilon_j) \neq 0
\end{equation}
which follows since all eigenvalues are \textit{pairwise distinct}. This implies that the matrix is invertible and only the trivial solution, $\tilde{a}_n^P = 0$ for all $n\in\mathbb{N}_0$, 
solves the system of equations in \eqref{VMEQS}. Consequently, we have proven that the eigenstates, which appear with nonzero probability in the corresponding measurement of the quantum state 
$|\phi\rangle$ and  which belong to pairwise distinct eigenvalues of the observable $\hat{\mathrm{H}}$ that commutes with all elements $\hat{P}\in\mathcal{P}$ are fully (anti)-symmetric 
if and only if the state $|\phi\rangle$ is fully (anti)-symmetric. 

\end{document}